\documentstyle[preprint,aps,epsf]{revtex}
\begin{document}
\draft

\title{\bf Percolation of strings and the first RHIC data on multiplicity
and transverse momentum distributions}
 
\author{M. A. Braun}
\address{High-energy physics department\\
St. Petersburg University \\
198904 St. Petersburg, Russia\\
}
\author{F. del Moral and C. Pajares}
\address{Departamento de F\'{\i}sica de Part\'{\i}culas\\
Universidade de Santiago de Compostela\\
15706 Santiago de Compostela, Spain
}

\date{\today}

\maketitle

\begin{abstract}
The dependence of the multiplicity on the number of
collisions and the transverse momentum distribution for central and
peripheral Au-Au collisions are studied in the model of percolation of 
strings relative to the experimental conditions  at RHIC.
The comparison with the first RHIC data
shows a good agreement. 

\end{abstract}


The first experimental data from the relativistic heavy ion collider show
several interesting features \cite{ref1,ref2,ref3,ref4,ref5,ref6,ref7}.
The measured multiplicity lies in the lower range of the values  
predicted by various models  and shows
a smooth dependence on the number of participants \cite{ref5,ref6}.
 The experimental $p_T$ distributions in Au-Au central collisions reveal a 
large departure
from the naive perturbative quantum cromodynamics (pQCD) predictions
 \cite{ref2,ref7,ref8}. Also, some differences are seen for 
peripheral collisions. In this paper, we compare these experimental results 
with the predictions deduced from the percolation of strings.

Multiparticle production is currently described in terms of color strings
stretched between the projectile and target, which decay into new strings 
and subsequently hadronize to produce observed hadrons. 
Color strings may be viewed as 
small areas in the transverse space, $\pi r_0^2$, $r_0$=0.2 fm, filled
with color field created by the colliding partons
 \cite{ref9,ref15}. Particles are produced via
emision of $q \bar q$ pairs in this field. With growing energy and/or atomic
number of colliding particles, the number of strings grows, and they start
to overlap, forming clusters, very much like disks in the two-dimensional
percolation theory \cite{ref9,ref10}. At a certain critical density a 
macroscopic cluster appears which marks the percolation phase transition.
The influence of the clustering of strings and their percolation on the
transverse momentum spectra was studied in \cite{ref13} in the 
thermodynamical limit
of a very large interaction area $S$ and very large number $N$ of formed 
strings, corresponding to very high energies and atomic numbers of 
participants. In this limit the study allows for an analytical treatment.
Here we apply this approach to realistic energies and colliding nuclei at
RHIC using Monte-Carlo simulations.

The percolation theory governs the geometrical pattern of the string 
clustering. Its observable implications, however, require introduction 
of some dynamics to describe string interaction, i.e., the behaviour of
a cluster formed by several overlapping strings.

There are several possibilities \cite{ref11,ref12,ref13}. Here we assume that
 a cluster
behaves as a single string with a higher color field $\vec Q_n$ corresponding
to the vectorial sum of the color charge of each individual $\vec Q_1$
string. The resulting color field covers the area $S_n$ of the cluster. As
$\vec Q_n=\sum_1^n \vec Q_1$, and the individual string colours may 
be oriented in an arbitrary manner respective to one another, the 
average $\vec Q_{1i}\vec Q_{1j}$ is zero, and $\vec Q_n^2=n\vec Q_1^2$.

Knowing this charge color $\vec Q_n$, one can compute the particle spectra 
produced by a single color string of area $S_n$ using the Schwinger
formula \cite{ref14}. For the multiplicity $\mu_n$ and
the average $p_T^2$ of particles, $<p_T^2>_n$, produced by a cluster
of $n$ strings one finds

\begin{equation} \mu_n=\sqrt{ n S_n \over S_1}\mu_1 \qquad ; \qquad  <p_T^2>_n
=\sqrt{ n S_1 \over S_n}<p_T^2>_1\label{ec1}\end{equation}
where $\mu_1$ and  $<p_T^2>_1$ are  the mean multiplicity and
$p_T^2$ of particles produced by a simple string with a transverse area 
$S_1=\pi r_0^2$. For strings just touching each other $S_n=nS_1$, and
hence $\mu_n=n\mu_1$, $<p_T^2>_n=<p_T^2>_1$, as expected. In the 
opposite case of 
maximum overlapping $S_n=S_1$ and therefore $\mu_n=\sqrt{n} \mu_1$, 
$<p_T^2>_n=\sqrt{n}<p_T^2>_1$, so that the multiplicity
results maximally supressed and the mean $p_T^2$ maximally 
enhanced. 
Notice that a  certain concervation rule holds for the total $<p_T^2>$:
$ \mu_n<p_T^2>_n=n\mu_1<p_T^2>_1$.

Eq. (\ref{ec1}) is the 
main tool of our calculation. In order to compute the
multiplicities we generate strings according to the quark-gluon
string model and using the Monte-Carlo code described in 
\cite{ref15}. Each string is produced at an identified impact parameter.
From this, knowing the transverse area of each string, we 
identify all the clusters formed in each collision and subsequently
compute for each of them its multiplicity in units
$\mu_1$. The value of $\mu_1$ was fixed by normalizing our results
to the SPS WA98 results for central Pb-Pb collisions \cite{ref16}.
Note that in colour string models both $\mu_1$ and $<p_T^2>$ are assumed
to rise with energy due to the increase of the rapidity interval 
available and hard scattering contribution. However this rise is very 
weak in the the energy range $\sqrt{s}=17\,\div\,200$ GeV. Therefore
we can neglect the energy dependence of both $\mu_1$ and $<p_T^2>_1$
in the first approximation.

\begin{figure}
  \centering\leavevmode
  \epsfxsize=5in\epsffile{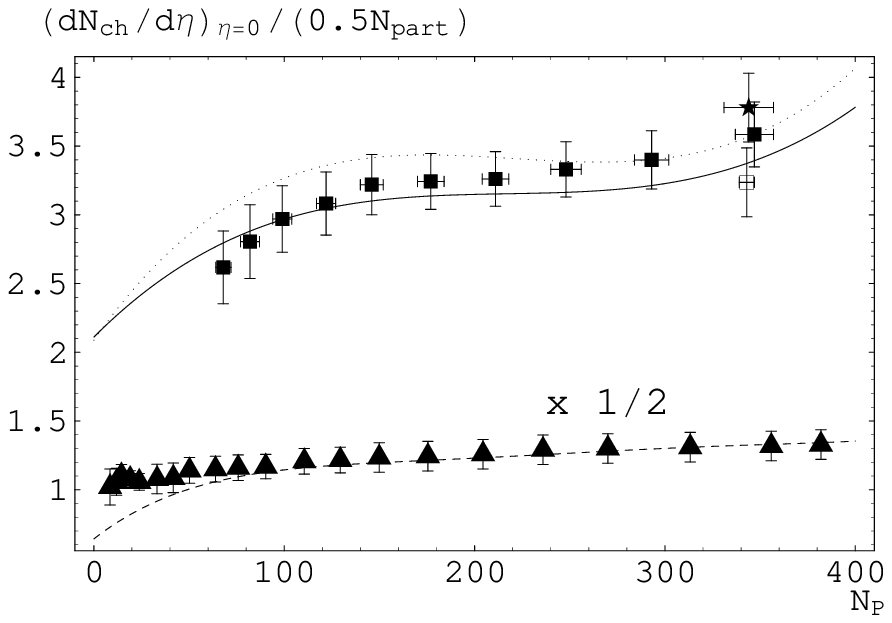}
\caption{Comparison of the 
dependence of the multiplicity on the number of participants
with the SPS WA98 [16]  data (filled triangles),  with the
RHIC PHENIX [5] (filled boxes) and PHOBOS [6] (non-filled boxes)
data at $\sqrt{s}=130$ GeV and with PHIC PHENIX data at $\sqrt{200}$
GeV (filled stars) . The dashed, solid and dotted lines are our 
predictions for the relevant energies.} \label{figure1}
\end{figure}

The comparison of our results for the dependence of the multiplicity
on the number of participants with the SPS WA98  data,
with the RHIC PHENIX \cite{ref5} and PHOBOS \cite{ref6}
data at $\sqrt{s}=130$ GeV and with PHOBOS data at $\sqrt{s}=200$ GeV is 
presented in Fig. \ref{figure1}. The agreement with the data at 
$\sqrt{s}=130$
GeV is quite good. Our predictions for $\sqrt{s}=200$ GeV lie a little below
the data. This could be a consequence of the mentioned weak energy 
dependence of $\mu_1$. On the whole the agreement with the data indicates 
that the percolation of strings mechanism correctly describes the 
behaviour of multiplicities in nucleus-nucleus collisions. This
conclusion has also been reached in  \cite{ref20}.

In order to compute the transverse momentum distributions, we make use of
the parametrization of the pp UA1 data at 130 GeV by A. Drees
\cite{ref8}
\begin{equation} {dN \over dp_t^2}= {a \over (p_0 + p_T)^\alpha}
\label{ec3}\end{equation}
where $a$,$p_0$ and $\alpha$ are parameters fitted to data \cite{ref2}. The
standardly expected  distribution for central Au-Au collision
is 
\begin{equation} {dN \over dp_T^2}\Big |_{AuAu} =   
<N_{\text{bin coll}}> {dN \over dp_T^2}\Big |_{pp} \label{ec4}\end{equation}
However, if we take into account percolation of strings then from Eq. 
\ref{ec1} we conclude
\begin{equation} <p_T^2>_{Au-Au}= <{nS_1 \over S_n}>_{Au-Au}^{1 / 2} 
<p_T^2>_1\label{ec5}\end{equation}
where, by definition,
\begin{equation} <{nS_1 \over S_n}>_{Au-Au}^{1 / 2}=
\frac{\sum ({nS_1 \over S_n})^{1/2}\mu_n}{\sum\mu_n} 
\label{ec51}\end{equation}
Here the sums go over all clusters formed in Au-Au collisions and $\mu_n$ 
are given by Eq. (1). To exclude the unknown $<p_T^2>_1$ we compare Eq. (4)
with the one for pp collisions
\begin{equation} <p_T^2>_{pp}= <{nS_1 \over S_n}>_{pp}^{1 / 2} 
<p_T^2>_1\label{ec52}\end{equation}
to finally get
\begin{equation}
<p_T^2>_{Au-Au}=<p_T^2>_{pp}\frac{<nS_1/S_n>_{Au-Au}^{1/2}}
{<nS_1/S_n>_{pp}^{1/2}}
\end{equation}
This equation implies that the same 
parametrization (\ref{ec3}) for the transverse momentum distribution
can be used for nucleus-nucleus collisions, with the only change
\begin{equation} p_0 \rightarrow p_0\left(\frac{ < nS_1/ S_n>_{Au-Au}}
{< nS_1/ S_n>_{pp}}\right)^{1/4}
\label{ec6}\end{equation}
Note that at the energies considered 
only two strings are exchanged in pp collisions
 on the average, with very 
small fusion probablilty, so that $< {nS_1/ S_n}>_{pp}\simeq 1$

\begin{figure}
  \centering\leavevmode
  \epsfxsize=5in\epsffile{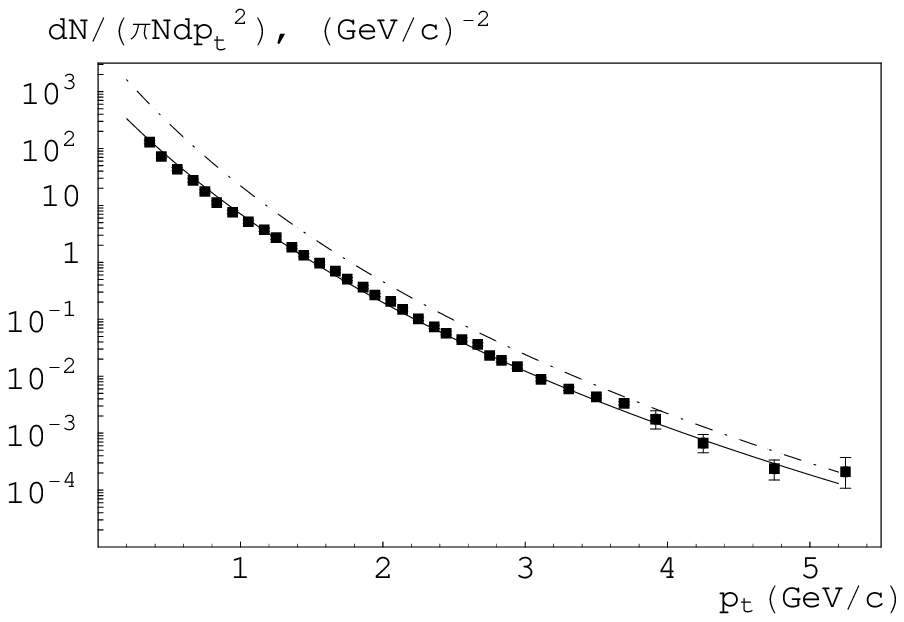}
\caption{Expected $p_T$ distribution using the percolation of string model
(solid line) for central (5\%) Au-Au collisions compared with PHENIX
experimental data [2,17] (filled boxes). Also it is shown
the expected distribution [8] given  by formula (\ref{ec4}) and (\ref{ec3}) (dotted-dashed
line).}
\label{figure2}
\end{figure}

\begin{figure}
  \centering\leavevmode
  \epsfxsize=5in\epsffile{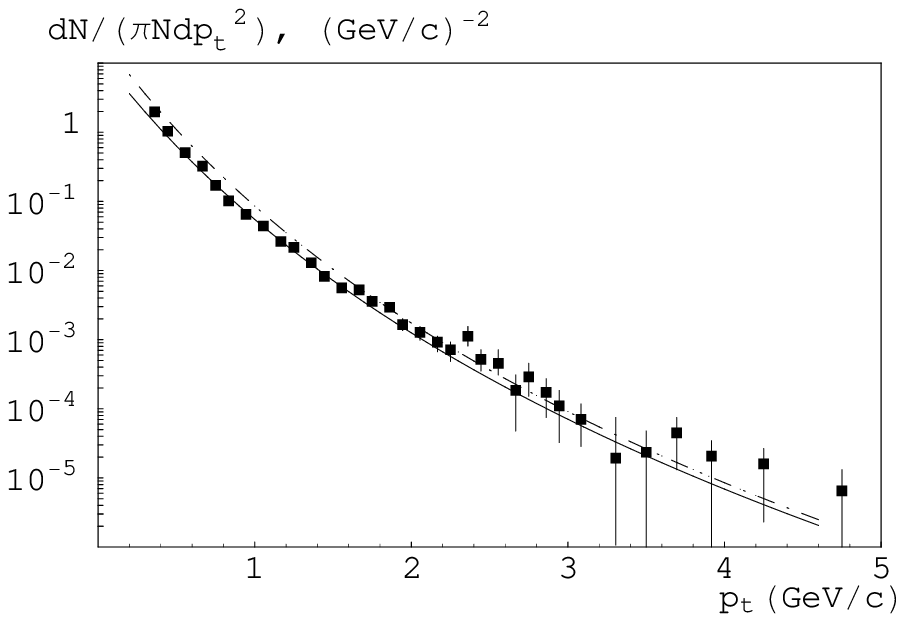}
\caption{Expected $p_T$ distribution using the percolation of string model
(solid line) for peripheral (80-90\%) Au-Au collisions compared with PHENIX
experimental data [2,17] (filled boxes). Also it is shown
the expected distribution [8] given  by formula (\ref{ec4}) and (\ref{ec3}) (dotted-dashed
line).}
\label{figure3}
\end{figure}

In Fig. \ref{figure2}, we show the 
distribution
(\ref{ec3}) with the change (\ref{ec6}) for central (5\%) Au-Au 
collisions (solid line) compared
to the PHENIX experimental data \cite{ref2,ref17} (black squares). Also 
the distribution expected from the independent string picture (Eq. 
(3)) is shown  (dash-dotted line).
For peripheral (80-92\%) Au-Au collisions the same results are presented in 
Fig. \ref{figure3} (with the same notations).
A very good agreement is observed in both cases. We stress that this 
result  is obtained 
using a very simple formula deduced in a simple way from the dynamics,
with practically a single parameter, the string transverse radius.

\begin{figure}
  \centering\leavevmode
  \epsfxsize=5in\epsffile{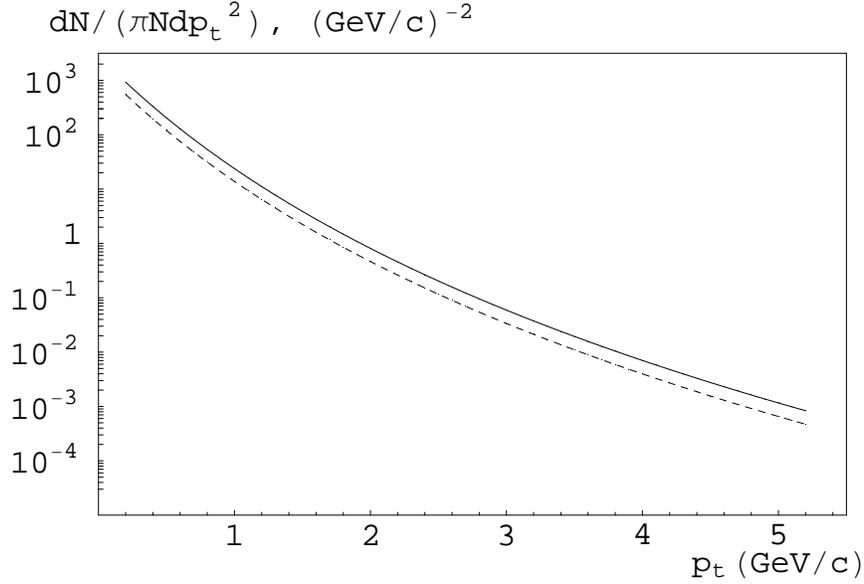}
\caption{Predictions for central (5\%) Au-Au (solid line) and central (5\%) Ag-Ag (dashed line)
collisions at 200 GeV/n.}
\label{figure4}
\end{figure}

Our predictions for central (5\%) Ag-Ag and central Au-Au collisions at
200 GeV/n are presented in Fig. \ref{figure4}.

In the thermodynamic limit one approximately obtains \cite{ref12} 
\begin{equation}
<\frac{nS_1}{S_n}>=\frac{\eta}{1-\exp(-\eta)}\equiv1/F^2(\eta)
\end{equation}
where $\eta=N\pi r_0^2/S$ is the percolation parameter assumed to be
finite with both $N$ and $S$ large. 
Then one obtains an analytic expression for the transverse momentum
distribution for all colliding particles in terms of 
$\eta$ 
\begin{equation} {dN \over dp_t^2}= 
{a \over\Big (p_0\sqrt{ F(\eta_{pp})/F(\eta)}  + p_T\Big)^\alpha}
\end{equation}
In (10) $a$ and $p_0$ are parameters of the experimentally observed spetrum 
in pp collisions at a given energy and $\eta$ and $\eta_{pp}$ are 
percolation parameters for nucleus-nucleus and pp collisions respectively,
to be obtained from the  geometrical percolation picture. Naturally all 
parameters depend on energy, although, as mentioned, at energies below 
200 GeV $\eta_{pp}$ is quite small so that $F(\eta_{pp})\simeq 1$.
Eq. (10) implies a certain universality of the distributions at fixed energy,
for all participant nuclei and different centralities: the distribution
results a universal function of $\eta$.  

At very high energies both $\eta$ and $\eta_{pp}$ become large,
so that $F(\eta)\simeq 1/\sqrt{\eta}$. Then we find
\begin{equation}
\frac{F(\eta_{pp})}{F(\eta)}\simeq\sqrt{\frac{\eta}{\eta_{pp}}}
\end{equation}
For minimum bias collision of identical nuclei
\begin{equation}
\eta=16\frac{\nu N_{pp}\pi r_0^2}{\sigma_{AA}^{tot}}
\end{equation}
where $\nu$ the average number of collisions and we used the fact that the
average overlap area is 1/16 of the total nucleus-nucleus cross-section.
From this we find an asymptotic expression for the distribution
valid at very high energies (large $\eta$'s):
\begin{equation} {dN \over dp_t^2}=
{a \over (2p_0 \sqrt{A\sigma_{pp}/\sigma^{tot}_{AA}}  + 
p_T)^\alpha}\end{equation}
Accordingly we find for these energies
\begin {equation}
<p_T^2>_{AA}=4<p_T^2>_{pp}\frac{A\sigma_{pp}}{\sigma_{AA}^{tot}}
\end{equation}
from which it evidently follows that at very high energies the ratio of the
averages  $p_t^2$ in nucleus-nucleus and pp collisions should grow
as the total pp cross-section.

Notice that from (1) we have for a cluster
\begin{equation}
<p_T^2>_n=\frac{S_1}{S_n}\frac{<p_T^2>_1}{\mu_1}\mu_n
\end {equation}
In the limit of very high density of strings, as observed from the 
Monte-Carlo calculations, practically all strings overlap into a single 
cluster, which occupies the whole interaction area. This is obviously
an alternative description of parton saturation. In this limit from (15)
we find a universal relation between the observed $<p_T^2>_{AB}$ and the 
multiplicity
per unit overlap area in the (nucleus A)-(nucleus B) collisions,
valid for any participant nuclei and any centrality:
\begin{equation}
<p_T^2>_{AB}=\frac{S_1<p_T^2>_1}{\mu_1}\frac{1}{S_{AB}}\mu_{AB}
\end {equation}
Here $S_{AB}$ is the overlap area depending on the impact parameter.
It is remarkable that a similar relation has been found
using the idea of the saturation of gluons in [25]. In their case
the proportionality coefficient is given by $\alpha_sP^2(m_h/p_s)$,
where $m_h$ is the mass of the produced hadron  $p_s$ is essentially the 
saturation momentum and $P$ is some function which tends to a constant as 
its argument vanishes. At high energies and therefore large $p_s$
this coefficient tends to a constant modulo logarithmic terms coming
from the running of the coupling constant. In our case the coefficient
also tends essentially to a constant at high energies, since both
$S_1<p_T^2>$ and $\mu_1$ (per unit rapidity) tend to a constant.

To conclude, some general remarks.
The colour string model, in principle, was introduced  to describe 
the soft part of the spectra.
However our results remain in good agreement with the experimental data 
up to $p_T$ of the order 5 GeV/c, where hard scattering is supposed to play
the dominant role, standardly desribed by the PQCD approach.
This implies that with the parametrization (\ref{ec3}) the hard part of the 
spectrum is also included into our string picture, corresponding to the
high-momentum tail of the string spectra.
Clustering and percolation of strings can then be looked upon as an 
alternative description of non-linear effects, which in the PQCD language
correspond to high partonic densities, such as  
saturation of partons.

There are several models which explain qualitatively the dependence of the
multiplicity on the number of participants 
\cite{ref18,ref19,ref20,ref21,ref22,ref23}, 
although 
not all of  them are able to reproduce quantitatively the experimental data.
The situation is worse concerning the $p_T$ distributions. 
With some fitted parameters, the jet
quenching effect \cite{ref8,ref24} could explain the data, but in this case
one expects an enhancement of the multiplicity in the central rapidity
region, which is not seen by experimental data. The percolation of 
strings is a natural and simple way of explaining both the
multiplicity and transverse momentum distribution. To further check our 
approach, experimental information on forward-backward correlations of 
multiplicity and transverse momentum distributions would be welcome
(see \cite{ref11,ref12}).

This work has been done under contract AEN99-0589-C02-02 from CICYT of
SPAIN and contract PGIDTOOPXI20613PN from Xunta de Galicia. It was also
supported by the Secretaria de Estado de Educacion y Universidades of Spain
and NATO grant PST.CLG.976799.

\end{document}